\documentclass{PoS}

\usepackage{amsthm, amssymb, amsmath, latexsym}

\usepackage{nicefrac}

\usepackage{graphicx}
\usepackage{subfigure}
\title{$\pi$-$\pi$ Scattering with $N_f=2+1+1$ Twisted Mass Fermions }
\ShortTitle{$\pi$-$\pi$ Scattering with $N_f=2+1+1$ Twisted Mass Fermions }

\author{Christopher Helmes, Christian Jost, \speaker{Bastian~Knippschild}, Liuming Liu, Carsten Urbach, Markus Werner\\\\
    HISKP (Theory), University of Bonn, Germany\\\\
	E-mail: \email{knippschild@hiskp.uni-bonn.de}\\\\}

\abstract{
\vspace{+0.2cm}

	$\pi$-$\pi$ scattering is investigated for the first time for $N_f=2+1+1$ dynamical quark flavours using Wilson twisted mass fermions. L{\"u}scher's finite size method is used to relate energy shifts in finite volume to scattering quantities like the scattering length in the I=2 channel. The computation is performed at several pion masses and lattice spacings utilising the stochastic LapH method.
	
}

\FullConference{The 32nd International Symposium on Lattice Field Theory,\\
		23-28 June, 2014\\
		Columbia University New York, NY}

\begin{document}

%%%%%%%%%%%%%%%%%%%%%%%%%%%%%%%%%
% 
% Introduction
% 
%%%%%%%%%%%%%%%%%%%%%%%%%%%%%%%%%
\section{Introduction}

Most particles in the hadron spectrum are resonances which are solely described by their mass and decay width. It appears that some of these resonance states like e.g. the roper resonance or scalar mesons below $2$ GeV are not even qualitatively described by the quark model. Therefore, a non-perturbative computation from first principles is highly desirable. Throughout this work lattice QCD is used which implements these requests.

A direct determination of scattering parameters on the lattice is not possible because of the Euclidean nature of lattice simulations. However, L{\"u}scher proposed a method to connect finite volume effects with scattering parameters \cite{Luscher:1986pf,Luscher:1990ux} which is used here. Due to finite volumes in lattice simulations the energy of a two particle system is shifted compared to twice the energy of a single particle. For most of the physically interesting channels this method requires one to precisely determine as many as possible energy levels in a given channel. This can be achived by applying the variational method to large correlation matrices.

A few years back a new method was introduced which combines an effective smearing technique and a stochastic approach, the so called stochastic Laplacian Heavyside smearing (sLapH) \cite{Peardon:2009gh,Morningstar:2011ka}. This method allows for an all-to-all approach in combination of an easy built-up of a large operator basis without the need to perform new and expensive inversions when changing the observable of interest. It is, hence, well suited for computing scattering properties, and it was successfully applied for mesons for instance in Refs.~\cite{Dudek:2010ew,Dudek:2012gj,Prelovsek:2014swa,Lang:2014yfa,Prelovsek:2013cra}. 
Here we extend on this by in particular focussing on three values of the lattice spacing and a wide range of pion masses.
For this purpose we use (yet a subset of) ensembles produced by the European twisted mass collaboration~\cite{Baron:2011sf,Baron:2010bv} with different pion masses, volumes, and lattice spacings as presented in tab.~\ref{table_configs}.  

In this proceedings we present preliminary results for $\pi$-$\pi$ scattering at isospin $I=2$. It represents a benchmark system for scattering processes. Other scattering processes are currently under investigation like the other isospin channels of $\pi$-$\pi$ scattering, $K$-$\pi$ scattering, and D-meson scattering \cite{Liuming}.

%%%%%%%%%%%%%%%%%%%%%%%%%%%%%%%%% 
% 
% Technical details
% 
%%%%%%%%%%%%%%%%%%%%%%%%%%%%%%%%% 
\section{Technical details}

For $I=2$, $\pi$-$\pi$ scattering the L{\"u}scher formula reads

\begin{equation}
  \delta E_{\pi\pi}^{I=2} = -\frac{4\pi a_{\pi\pi}^{I=2}}{m_\pi L_s^3}\left\{ 1+ c_1\frac{a_{\pi\pi}^{I=2}}{L_s} + 			c_2\frac{\left(a_{\pi\pi}^{I=2}\right)^2}{L_s^2} \right\} + \mathcal{O}(L_s^{-6}),
  \label{eq_energyshift}
\end{equation}

where $c_1$ and $c_2$ are known numerical constants~\cite{Luscher:1986pf} and $a_{\pi\pi}^{I=2}$ is the scattering length.

For computing $\delta E$ two- and four-point functions need to be computed. For the isospin-2 channel only connected contributions need to be taken into account. As mentioned before, we are using the sLapH method introduced in Ref.~\cite{Morningstar:2011ka}. We refer the reader to this reference for the details. We note here that we can re-use the inversion for this project for many other physically interesting channels without the need for new inversions.

For the operator construction Fiertz rearrangement has to be taken into account~\cite{Fukugita:1994ve}. Following Ref.~\cite{Fukugita:1994ve} we deal with this issue by placing the two pions in our two-pion operator at adjacent time slices.

\begin{table}[t!]
  \begin{center}
    \begin{tabular}{|c|c|c|c|c|c|c|c|c|c|}
      \hline
      name & A30 & A40.32 & A40.24 & A40.20 & D45 & B55 & A60 & A80 & A100 \\ \hline
      $L_s(L_t)$ & 32(64) & 32(64) & 24(48) & 20(48) & 32(64) & 32(64) & 24(48) & 24(48) & 24(48) \\ \hline
      $\nicefrac{m_\pi}{f_\pi}$  & 1.86 & 2.06(1) & 2.03(3) & 2.11(5) & 2.49 & 2.34 & 2.32 & 2.55 & 2.77 \\ \hline
      \# conf & 100 & 150 & 200 & 150 & 50 & 50 & 200 & 300 & 300 \\ \hline
    \end{tabular}
  \end{center}
  \caption{List of ensembles used in this work, including spatial volume, $L_s$, the temporal lattice extent, $L_t$, the ratio of pion masses and pion decay constant, $\nicefrac{m_\pi}{f_\pi}$, which is already corrected for volume effects, and the number of configurations. The lowest pion mass used in this work is $m_\pi=284$~MeV from ensemble A30.}
  \label{table_configs}
\end{table}

We have tuned the number of eigenvectors $N_{ev}$ for the sLapH method by minimising excited state contributions in simple pseudoscalar correlation functions. We found $N_{ev} = 66$ for $L=20$, $N_{ev} = 120$ for $L=24$ and, $N_{ev} = 220$ for $L=32$ to suppress excited states sufficiently. The eigenvectors are combined with random vectors to reduce the number of inversions. These random vectors carry temporal, Dirac, and eigenvector indices. Improvement of the signal-to-noise ratio can be achieved by combining the random vectors with a dilution procedure which can be applied in all three indices individually \cite{Morningstar:2011ka}. We chose full dilution in Dirac space and an interlace dilution in eigenvector space. In temporal coordinates we use a block dilution scheme. The size of the dilution matrices was tuned in a way that we end up with a total number of 512 or 572 inversions independent of the volume.

\begin{figure}[h]
\begin{minipage}[t]{0.45\textwidth}
	\centering
	\hspace*{-1.0cm}
	\includegraphics[width = 1.3\textwidth]{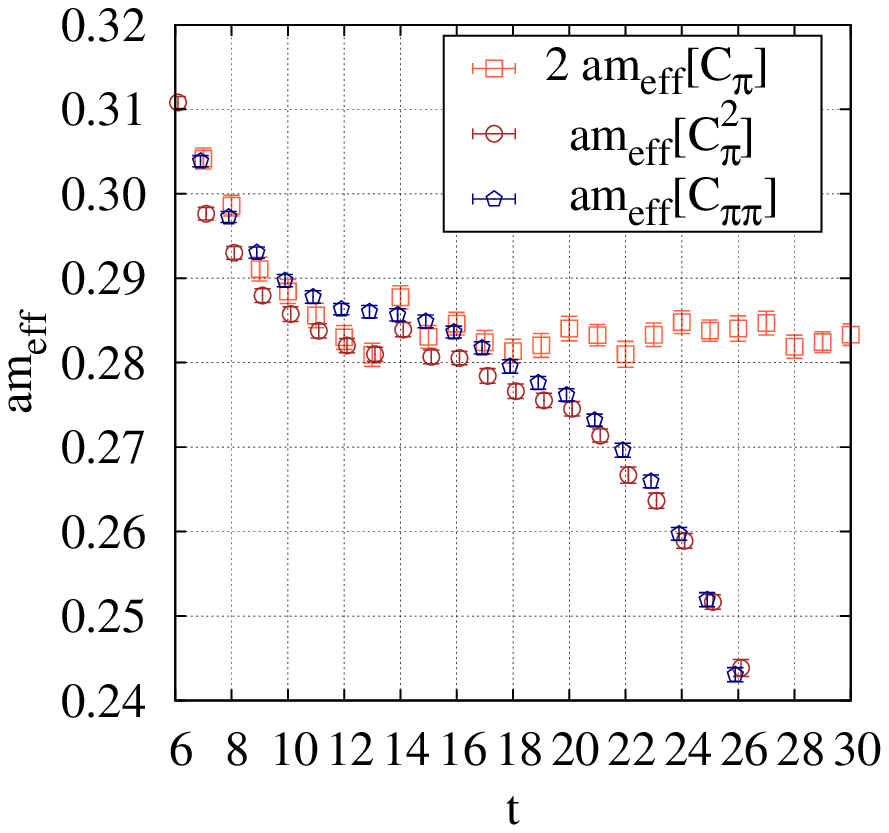}
	\vspace*{-0.75cm}
	\caption{Effective masses computed on A40.32 from the two-pion correlation function squared, $a m_\text{eff}\left[C^2_\pi\right]$, and from the four-point correlation function, $a m_\text{eff}\left[C_{\pi\pi}\right]$, compared to twice the effective mass of a single pion, $2a m_\text{eff}\left[C_\pi\right]$.}
	\label{fig_thermal_states}
\end{minipage}
\hfill
\begin{minipage}[t]{0.45\textwidth}
	\centering
	\hspace*{-1.0cm}
	\includegraphics[width = 1.3\textwidth]{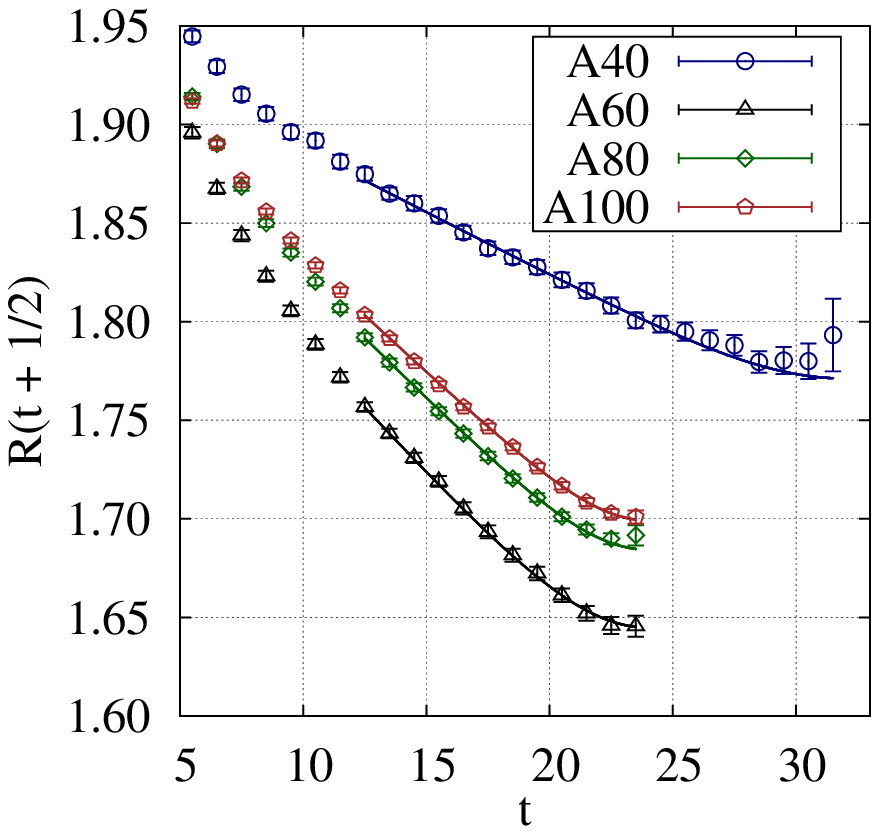}
	\vspace*{-0.75cm}
	\caption{Plot of $R(t+1/2)$ eq.~\protect\ref{eqnratio1} for four ensembles, see legend. The lines represent our best fits to the data.}
	\label{fig:ratio}
\end{minipage}
\end{figure}

Due to periodic boundary conditions so-called thermal states need to be taken into account, first discussed for $I=2$ pion scattering and twisted mass fermions in Ref.~\cite{Feng:2009ij}. For $\pi$-$\pi$ scattering the thermal states are constant contributions in Euclidean time and occur if the temporal lattice extend is not infinite. They can be observed in fig.~\ref{fig_thermal_states} where effective masses from three different pion correlation functions are compared. At early time slices all masses show contaminations from excited states. However, at large Euclidean times only the effective mass from a single pion correlation function, $2a m_\text{eff}\left[C_\pi\right]$, shows a plateau. The four-point function and squared two-point function do not show a plateau at all because of thermal state contaminations.

One way to deal with thermal states is to study the following ratio~\cite{Feng:2009ij}

\begin{align}
  R(t + \nicefrac{1}{2}) & = \frac{C_{\pi\pi}(t)-C_{\pi\pi}(t+1)}{C_{\pi}^2(t) - C_{\pi}^2(t+1)}\label{eqnratio1} \\[+2ex]
  & = A\left( \cosh(\delta E_{\pi\pi}^{I=2}t^\prime) + \sinh(\delta E_{\pi\pi}^{I=2}t^\prime) \coth(2m_\pi t^\prime)\right),
  \label{eqnratio2}
\end{align}

where $C_{\pi\pi}$ is the two-pion correlation function, $C_{\pi}^2$ is the squared single pion correlation function, $t$ is the Euclidean time and $t^\prime = t + \frac{1}{2} - \frac{T}{2}$. By taking differences the constant thermal state contributions cancel. The ratio, $R$, is shown together with our best fits to the data in fig.~\ref{fig:ratio}.

%%%%%%%%%%%%%%%%%%%%%%%%%%%%%%%%% 
% 
% Preliminary results
% 
%%%%%%%%%%%%%%%%%%%%%%%%%%%%%%%%% 
\section{Preliminary results}

\begin{figure}[h]
  \centering
  \includegraphics[width=0.6\textwidth]{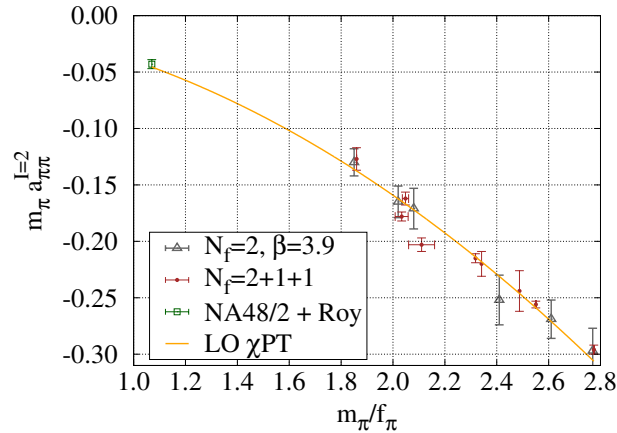}	
  \vspace{-0.5cm}
  \caption{Summary of our data for $m_\pi a_{\pi\pi}^{I=2}$. The green square shows the value extracted from experimental data~\cite{Batley:2000zz,BlochDevaux:2009zzb} and the grey triangles are data from the previous $N_f=2$ study~\cite{Feng:2009ij}. The line is the $\chi$PT prediction and our data are shown as red circles.}
  \label{fig_summary}
\end{figure}

Our preliminary results for the $I=2$ pion scattering length $m_\pi a_{\pi\pi}^{I=2}$ in dependence of $\nicefrac{m_\pi}{f_\pi}$ are summarised in fig.~\ref{fig_summary}. The green square marks the value extracted from experimental data using Roy equations~\cite{Batley:2000zz,BlochDevaux:2009zzb}. Our results are shown as red circles while the grey triangles mark the results from a previous $N_f=2$ study~\cite{Feng:2009ij} with twisted mass fermions. The line is the leading order $\chi$PT prediction \cite{Bijnens:1997vq}

\begin{equation}
  m_\pi a_{\pi\pi}^{I=2} = -\frac{m_\pi^2}{16\pi^2 f_\pi^2},
\end{equation}

which is not a fit but solely determined by the pion mass and its decay constant. All points, $N_f=2$ and $N_f=2+1+1$, do agree with each other within errors and with the $\chi$PT prediction. It must be noted that no correction for any systematic effects of the scattering length was included yet. The only correction applied was a correction for volume effects to $\nicefrac{m_\pi}{f_\pi}$ as determined in ref.~\cite{Carrasco:2014cwa}.

The $N_f=2$ data show larger statistical errors than our data which can be explained by the number of inversions used. While for the $N_f=2$ only four inversions per configuration were used, we used about 512. The number of configurations we used is about a factor of two to 10 smaller than what was used in ref.~\cite{Feng:2009ij}. However, sLapH is still much more expensive when only comparing the computation of the $I=2$ pion scattering length. The strength of sLapH is the reusability of propagators when changing to another observable. In the long run sLapH will be much more efficient than conventional methods.

%%%%%%%%%%%%%%%%%%%%%%%%%%%%%%%%% 
% 
% Systematic effects - an outlook
% 
%%%%%%%%%%%%%%%%%%%%%%%%%%%%%%%%% 
\section{Systematic effects - an outlook}

Systematic effects which can influence the extraction of $a_{\pi\pi}^{I=2}$ are discussed in great detail in Ref.~\cite{Feng:2009ij}. It was found that on the level of the statistical accuracy reached in Ref.~\cite{Feng:2009ij} the systematic uncertainties appeared to be negligible. However, since our statistical errors are smaller, we have to investigate these effects carefully in the future. Among these are possible effects from parity and isospin violation at non-zero lattice spacing in twisted mass lattice QCD, even though no evidence for these was found in the $N_f=2$ study~\cite{Feng:2009ij}.

Moreover, two types of volume effects need to be taken into account. The first one is exponentially suppressed by $\sim e^{-m_\pi L}$, and comes from the self interaction of a single pion in a finite box~\cite{Luscher:1985dn}. This effect can be corrected with the help of $\chi$PT via

\begin{equation}
  (m_\pi a_{\pi\pi}^{I=2})_L = (m_\pi a_{\pi\pi}^{I=2})_\infty + \Delta_\text{FV}(m_\pi, L) \nonumber
\end{equation}

where $\Delta_\text{FV}(m_\pi, L)$ is known~\cite{Bedaque:2006yi}. We find that this effect can already now be as large as $50\%$ of our statistical error. We remark that an analysis of these finite volume effects in twisted mass Wilson $\chi$PT similar to Ref.~\cite{Bar:2010jk} would be useful.

\begin{figure}[h]
  \centering
  \includegraphics[width=.6\textwidth]{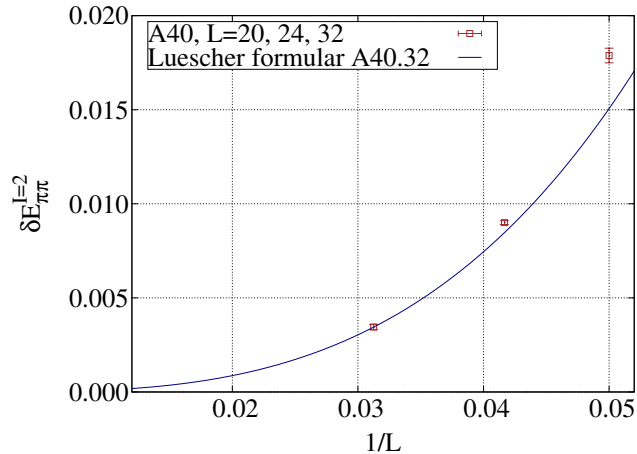}
  \vspace{-0.5cm}
  \caption{The energy shift in dependence of the volume is shown to illustrate volume effects. The blue line corresponds to the result taken from the largest volume.}
  \label{fig_volume_effects}
\end{figure}

The second and more important volume effect is only power suppressed in $\nicefrac{1}{L}$.  L{\"u}scher's formula, eq.~\ref{eq_energyshift}, is based on the effective range expansion (ERE) to lowest order

\begin{equation}
  \lim_{k\to 0} k \cot\left(\delta_{s}(k)\right) = -\frac{1}{a_{s}},
  \label{eqn_effectiverangefirstorder}
\end{equation}

where $k$ is the volume dependent scattering momentum and $\delta_{s}(k)$ is the scattering phase shift. For too small volumes, $k$ is not even close to zero and higher orders need to be taken into account. This is illustrated in fig.~\ref{fig_volume_effects} where the energy shift is plotted in dependence of $\nicefrac{1}{L_s}$ for the three A40 ensembles, which differ only in $L_s$ and  $L_t$===. The blue curve represents eq.~\ref{eq_energyshift} with $m_\pi a_{\pi\pi}^{I=2}$ computed from the largest volume ($L_s=32$). The curve misses the two other points for $L_s=24$ and $L_s=20$, indicating additional volume effects at least for these volumes at this pion mass.

A possible way to deal with nonzero $k$ would be use the ERE to higher order 

\begin{equation}
  \frac{k \cot{\left(\delta_{\pi\pi}^{I=2}(k)\right)}}{m_\pi} = -\frac{1}{m_\pi a_{\pi\pi}^{I=2}} + \frac{1}{2}m_\pi r_\text{reff}\frac{k^2}{m_\pi^2} + \mathcal{O}(k^4), 
  \label{eq_effectiverangeexpension}
\end{equation}

where the effective range, $r_\text{reff}$, appears. This would enter eq.~\ref{eq_energyshift} at $\mathcal{O}(L^{-6})$ which would lead to one equation with two unknowns, the scattering length and the effective range. The better solution to this problem is to compute $k$ directly from the energy shift via a dispersion relation, see e.g. Ref.~\cite{Beane:2007xs}. With $k$, the expression $\frac{k \cot{\left(\delta_{\pi\pi}^{I=2}(k)\right)}}{m_\pi}$ can be computed via L{\"u}scher's Zeta function \cite{Luscher:1986pf}. Once the phase shift is known as a function of $k^2$, other quantities like the scattering length can be computed. For the ensembles for which we have only one volume available we are going to use different momenta and moving frames for this purpose. For a detailed review of this method see e.g. Refs.~\cite{Beane:2011sc,Dudek:2012gj}.

%%%%%%%%%%%%%%%%%%%%%%%%%%%%%%%%% 
% 
% Conclusions and Outlook
% 
%%%%%%%%%%%%%%%%%%%%%%%%%%%%%%%%% 
\section{Conclusions and outlook}

We made a first attempt to address scattering processes in $N_f=2+1+1$ twisted mass lattice QCD with the stochastic LapH method. The focus was on the isospin 2 $\pi$-$\pi$ scattering length as a first test of our methods. We find good agreement with the $N_f=2$ results~\cite{Feng:2009ij}, however investigation of systematic effects might become demanding because of small statistical errors. Larger lattices with $L=48$ and smaller pion masses down to the physical point are on its way to extend the results presented here.
In the future we will use the stochastic LapH method to build a larger operator basis with momenta and displacements to be able to address more scattering processes of more exotic particles.

%%%%%%%%%%%%%%%%%%%%%%%%%%%%%%%%% 
% 
% acknowledgement
% 
%%%%%%%%%%%%%%%%%%%%%%%%%%%%%%%%% 
\section{Acknowledgements}

This project was funded by the DFG as a project in the Sino-German CRC 110. The computer time for this project was made available to us by the John von Neumann-Institute for Computing (NIC) and JSC on the Juqueen system in J{\"u}lich and on the DFG funded GPU cluster QBiG in Bonn.

\newpage
%%%%%%%%%%%%%%%%%%%%%%%%%%%%%%%%% 
% 
% references
% 
%%%%%%%%%%%%%%%%%%%%%%%%%%%%%%%%% 
\bibliographystyle{h-elsevier}
\bibliography{Refs}

\end{document}